\documentclass[10pt]{article}

\usepackage{amsmath}
\usepackage{amssymb}

\usepackage{cite}

\usepackage{hyperref}

\usepackage{lineno}

\usepackage{microtype}
\DisableLigatures[f]{encoding = *, family = * }
\usepackage{graphicx}
\usepackage{float}

\usepackage[labelfont=bf,labelsep=period,justification=raggedright]{caption}

\makeatletter
\renewcommand{\@biblabel}[1]{\quad#1.}
\makeatother

\date{}

\begin{document}

\begin{flushleft}
{\Large
\textbf{How volatilities nonlocal in time affect the price dynamics in complex financial systems}
}
\\
Lei Tan\textsuperscript{1,2},
Bo Zheng\textsuperscript{1,2,*},
Jun-Jie Chen\textsuperscript{1,2},
Xiong-Fei Jiang\textsuperscript{1},
\\
\bf{1} Department of Physics, Zhejiang University, Hangzhou 310027, China
\\
\bf{2} Collaborative Innovation Center of Advanced Microstructures, Nanjing University, Nanjing 210093, China
\\
* E-mail: zhengbo@zju.edu.cn
\end{flushleft}

\section*{Abstract}
What is the dominating mechanism of the price dynamics in financial systems is of great interest to scientists.
The problem whether and how volatilities affect the price movement draws much attention. Although many efforts have been made,
it remains challenging. Physicists usually apply the concepts and methods in statistical physics,
such as temporal correlation functions, to study financial dynamics. However, the usual volatility-return correlation function, which is local in time,
typically fluctuates around zero. Here we construct dynamic observables nonlocal in time to explore the volatility-return
correlation, based on the empirical data of hundreds of individual stocks and 25 stock market indices in different countries. Strikingly, the correlation
is discovered to be non-zero, with an amplitude of a few percent and a duration of over two weeks. This result provides compelling evidence
that past volatilities nonlocal in time affect future returns. Further, we introduce an agent-based model with a novel mechanism, that is,
the asymmetric trading preference in volatile and stable markets, to understand the microscopic origin of
the volatility-return correlation nonlocal in time.
\section*{Author Summary}

\section*{Introduction}
Financial markets, as a kind of typical complex systems with many-body interactions, have drawn much attention of scientists.
In recent years, for example, various concepts and methods in statistical physics have been applied and much progress has been achieved\cite{man95,ple99,gop99,bou01,kra02,gab03,onn03,sor03,qiu06,pod09,she09,ken11,pre11,sha11,fen12,ken12,pre12,che13,ken13,jia14,maj14,yur14}.
Following the trend towards quantitative analysis in finance, the efforts of scientists in different fields promote each other and deepen our understanding
of financial systems\cite{bla76,fre87,cam92,man95,bek00,con00,gab03,yam05,bol06,osi07,sha09,ken10a,ken10b,ren10,da11,zha11,pre12,jia12}.

From the perspective of physicists, a financial market is regarded as a dynamic system, and the price dynamics,
i.e. the time evolution of stock prices, can be characterized by temporal correlation functions, which describe
how one variable statistically changes with another.
It is well-known that the price volatilities are long-range correlated in time, which is called volatility clustering. Many activities have been
devoted to the study of the collective behaviors related to volatility clustering in stock markets\cite{gop99,liu99,egu00,kra02,gab03}.
However, our understanding on the movement of the price return itself is very much limited. The autocorrelating time of returns is extremely short,
that is, on the order of minutes\cite{gop99,liu99}. As to higher-order time correlations, it is discovered that the return-volatility correlation is negative
--- in other words, past negative returns enhance future volatilities\cite{bla76,cox76,bou01,qiu06,she09a}. This is the so-called
leverage effect. As far as we know, all stock markets in the world exhibit the leverage effect except for the Chinese stock market, which
unexpectedly shows an anti-leverage effect, i.e., the correlation between past returns and future volatilities is positive\cite{qiu06,she09a}.
Returns represent the
price changes, and volatilities measure the fluctuations of the price movement. The leverage and anti-leverage effects characterize how
price changes induce fluctuations. At this stage, one may ask what affects the
return itself. It has been discovered that future returns can be predicted by the dividend-price ratio\cite{cam88,fam88},
which is corroborated by subsequent studies. However, the predictive power of the dividend-price ratio is sensitive to the selection
of the sample period\cite{val03,bou07}. Recently, price extrema are found to be linked with peaks in the volume time series\cite{pre11}.
Moreover, it is reported that
massive data sources, such as Google Trends and Wikipedia, contain early signs
of market moves. The argument is that these ``big data'' capture investors' attempts to gather information before decisions are
made\cite{da11,moa13,pre13}. These researches provide insight into the price dynamics.

What is the dominating mechanism of the price dynamics is highly complicated. The problem how volatilities
affect the price dynamics has drawn much attention.
Although many efforts have been made, it remains enormously challenging. According to a hypothesis known as the volatility feedback effect,
an anticipated increase in volatility would raise the required return in the future. To allow for higher future returns, the current
stock price decreases\cite{fre87,cam92}. Based on this hypothesis, various models, such as Generalized AutoRegressive
Conditional Heteroskedasticity (GARCH) model\cite{bol86} and Exponential GARCH (EGARCH) model\cite{nel91}
have been applied to examine the correlation between past volatilities and future returns, and the results are controversial.
The correlation is discovered to be positive in some researches\cite{fre87,cam92}, while negative in others\cite{tur89,nel91,glo93}. Often the coefficient
linking past volatilities to future returns is statistically insignificant\cite{bek00}. On the other hand, the volatility-return correlation function can be used
to characterize the correlation between past volatilities and future returns. If the hypothesis of the volatility feedback effect is valid, the volatility-return
correlation function should be non-zero. However, it typically fluctuates around zero\cite{bou01,qiu06}. Such a
volatility-return correlation function can only characterize the correlation local in time. In fact, the scenario in financial markets may be more
complicated. Interactions, and thus correlations could be nonlocal in time.

In this study, we construct a class of dynamic observables nonlocal in time to explore the volatility-return correlation, based on the empirical data of
hundreds of individual stocks in the New York and Shanghai stock exchanges, as well as $25$ stock market indices in different countries. Strikingly, the correlation is
discovered to be non-zero, with an amplitude of a few percent and a duration over two weeks. This result provides compelling
evidence that past volatilities nonlocal in time affect future returns. Further, we introduce an agent-based model with a novel mechanism,
that is, the asymmetric trading preference in volatile and stable markets, to understand the microscopic origin of the volatility-return correlation nonlocal in time.

\section*{Materials}
We collect the daily closing prices of $200$ individual stocks in the New York Stock Exchange (NYSE), $200$ individual
stocks in the Shanghai Stock Exchange (SSE)
and $25$ stock market indices in different countries. The time periods of the individual stocks and stock market indices are presented in Table~\ref{t1}.
All these data are obtained from Yahoo! Finance (finance.yahoo.com).
To keep the time periods for all
stocks exactly the same and as long as possible, we select $200$ stocks
in the SSE, most of which are large-cap stocks. For comparison, $200$ stocks
in the NYSE are collected.

\begin{table}[H]
 \centering
\caption{\textbf{The time period, effective pair of time windows and maximum $AP_{0}$.} The time period, effective pair of $T_{1}$ and $T_{2}$,
and maximum $AP_{0}$ for the individual stocks in the NYSE and SSE, as well as $18$ stock indices. The volatility-return correlation nonlocal in time is positive
for all these indices and stocks, except for the Australia and Japan indices, which exhibit a negative volatility-return correlation. For other $7$ indices, nonzero
$\Delta P(t)$ could not be detected for almost all pairs of $T_{1}$ and $T_{2}$. These indices include MERV (Argentina 1996-2012), BSESN (India 1997-2012),
KLSE (Malaysia 1993-2012), KJSE (Indonesia 1997-2011), OMXC20.CO (Denmark 2000-2012), OSEAX (Norway 2001-2012) and FTSE (England 1984-2012),
which are not listed in this table.}
\label{t1}
\begin{tabular}{ccccc}
\hline
Index & Period & Effective $T_{1}$ & Effective $T_{2}$ & max $AP_{0}$\tabularnewline
\hline
200 stocks in the NYSE & 1990-2006 & 6-36 & 45-250 & 0.006\tabularnewline
200 stocks in the SSE & 1997-2007 & 4-44 & 95-250 & 0.032\tabularnewline
\hline
BVSP (Brazil) & 1993-2012 & 26-44 & 60-105 & 0.027\tabularnewline
GSPTSE (Canada) & 1977-2012 & 19-40 & 190-240 & 0.029\tabularnewline
IPSA (Chile) & 2003-2012 & 31-44 & 190-220 & 0.032\tabularnewline
Shanghai Index (China) & 1990-2009 & 27-44 & 80-105 & 0.036\tabularnewline
S\&P 500 (America) & 1950-2011 & 29-36 & 185-225 & 0.012\tabularnewline
DAX (German) & 1959-2009 & 28-44 & 85-250 & 0.015\tabularnewline
KOSPI (Korea) & 1997-2012 & 26-44 & 90-115 & 0.027\tabularnewline
MXX (Mexico) & 1991-2012 & 6-19 & 65-140 & 0.029\tabularnewline
NZ50 (New Zealand) & 2004-2012 & 27-32 & 100-120 & 0.033\tabularnewline
IBEX (Spanish) & 1993-2012 & 11-25 & 105-225 & 0.026\tabularnewline
OMX (Sweden) & 1998-2012 & 7-17 & 55-160 & 0.043\tabularnewline
SSMI (Switzerland) & 1990-2012 & 27-44 & 80-110 & 0.023\tabularnewline
FCHI (France) & 1990-2012 & 18-19 & 135-145 & 0.021\tabularnewline
AEX (Holland) & 1982-2012 & 21-23 & 50-65 & 0.023\tabularnewline
Shenzhen (China) & 1991-2009 & 3-31 & 90-250 & 0.045\tabularnewline
DJI (America) & 1928-2011 & 39-41 & 200-225 & 0.006\tabularnewline
AORD (Australia) & 1984-2012 & 11-44 & 45-250 & -0.032\tabularnewline
N225 (Japan) & 1984-2011 & 34-44 & 200-250 & -0.017\tabularnewline
\hline
\end{tabular}
\end{table}

\section*{Methods and Results}
\subsection*{Asymmetric conditional probability in volatile and stable markets}
To explore the volatility-return correlation in stock markets, we construct a class of observables, including conditional
probabilities and correlation functions. We first discuss the conditional probabilities.

The price of a financial index or individual stock at time $t'$ is denoted by $Y\left(t'\right)$, and the logarithmic return is defined
as $R\left(t'\right)\equiv\ln Y\left(t'\right)-\ln Y\left(t'-1\right)$. For comparison
of different indices or stocks, we introduce the normalized return
\begin{equation}
r\left(t'\right)\equiv\left[R\left(t'\right)-\left\langle R\left(t'\right)\right\rangle \right]/ \sigma.
\label{e10}
\end{equation}
Here $\left\langle \cdots\right\rangle $ represents the average over time $t'$. In other words,
$\left\langle R\left(t'\right)\right\rangle=\left[\sum_{i=1}^n R(i)\right]/n$ is the average of the time series $R\left(t'\right)$, where $n$ denotes the total number of
the data points of $R\left(t'\right)$, and $\sigma=[\left\langle R^{2}\right\rangle -\left\langle R\right\rangle ^{2}]^{1/2}$ is the standard deviation of $R\left(t'\right)$.
There may be various definitions of volatility, a simplified one is
\begin{equation}
v\left(t'\right)=\left|r\left(t'\right)\right|,
\label{e20}
\end{equation}
which measures the magnitude of the price fluctuation.

One may compute temporal correlation functions
to investigate the dynamic correlations. The usual volatility-return correlation function is defined as $f(t)=\langle v\left(t'\right)\cdot r\left(t'+t\right)\rangle$
with $t>0$, and it characterizes how the volatility at  $t'$ influences the return at $t'+t$. However, this correlation function fluctuates around
zero\cite{bou01,qiu06}.
It is noteworthy that such a kind of $f(t)$ is local in time, while interactions such as information exchanges in financial markets may be more complicated,
leading to correlations nonlocal in time.

To explore the correlations nonlocal in time, we first define an average volatility at $t'$ over a past period of time $T$,
\begin{equation}
\left\langle v\left(t'\right)\right\rangle _{T}=\frac{1}{T}{\sum_{i=1}^T}v\left(t'-i+1\right).
\label{e30}
\end{equation}

To evaluate whether the average fluctuation in a short time period $T_{1}$ is strong or weak, we compare it with a background fluctuation,
which is defined over a much longer period of time $T_{2}$ in the past. Therefore, we introduce the difference of the average volatilities in two
different time windows,
\begin{equation}
\Delta v\left(t'\right)=\left\langle v\left(t'\right)\right\rangle _{T_{1}}-\left\langle v\left(t'\right)\right\rangle _{T_{2}},
\label{e40}
\end{equation}
with $T_{2} \gg T_{1}$. $T_{1}$ and $T_{2}$ are called the short window and long window, respectively. When $\Delta v\left(t'\right)>0$, the stock
market in the time window $T_{1}$ is volatile; otherwise, it is relatively stable.

Next, we compute the conditional probability $P^+(t)|_{\Delta v(t')>0}$, which is the probability of $r(t'+t)>0$ on the condition of $\Delta v(t')>0$.
Here we consider only $t>0$. Correspondingly, the conditional probability $P^+(t)|_{\Delta v(t')<0}$ is the probability of $r(t'+t)>0$
for $\Delta v(t')<0$. We do not observe any $r(t'+t)$ equal to $0$ in the normalized return series. Thus, the conditional probability
of $r(t'+t)<0$ is $1-P^+(t)|_{\Delta v(t')>0}$
and $1-P^+(t)|_{\Delta v(t')<0}$, respectively. In a time series of returns, the total number of positive returns is generally
different from that of negative ones.
Let us denote the unconditional probability that the return is positive by $P_{0}(t)$, which is the percentage of the positive returns in all returns
without any condition.

The specific calculations for $P^+(t)|_{\Delta v(t')>0}$, $P^+(t)|_{\Delta v(t')<0}$ and $P_{0}(t)$ are described in \nameref{S1_Appendix}.
If past volatilities and future returns do not correlate with each other, both $P^+(t)|_{\Delta v(t')>0}$ and $P^+(t)|_{\Delta v(t')<0}$ should be equal to $P_{0}(t)$.
In other words, if $P^+(t)|_{\Delta v(t')>0}$ and $P^+(t)|_{\Delta v(t')<0}$ are different from $P_{0}(t)$, i.e., if the conditional probability of returns
is asymmetric in volatile and stable markets, there exists a non-zero volatility-return correlation and such a correlation is nonlocal in time. In this case,
it can be proven that if $P^+(t)|_{\Delta v(t')>0}>P_{0}(t)$, we have $P^+(t)|_{\Delta v(t')<0}<P_{0}(t)$, otherwise,
we have $P^+(t)|_{\Delta v(t')<0}>P_{0}(t)$ (see \nameref{S1_Appendix}). To describe the asymmetric conditional probability in volatile and stable markets,
we introduce
\begin{equation}
\Delta P(t)=P^+(t)|_{\Delta v(t')>0}-P^+(t)|_{\Delta v(t')<0}.
\label{e50}
\end{equation}
It is important that the probability difference $\Delta P(t)$ relies on $\Delta v(t')$, thereby depending on the time windows $T_{1}$ and $T_{2}$.
Even though the volatility-return correlation function local in time is zero, the nonlocal observable $\Delta P(t)$ can be non-zero.
We call a pair of $T_{1}$ and $T_{2}$ at which $\Delta P(t)$ is non-zero an effective pair.

\begin{figure}[H]
 \centering
    \includegraphics[height=2.2in]{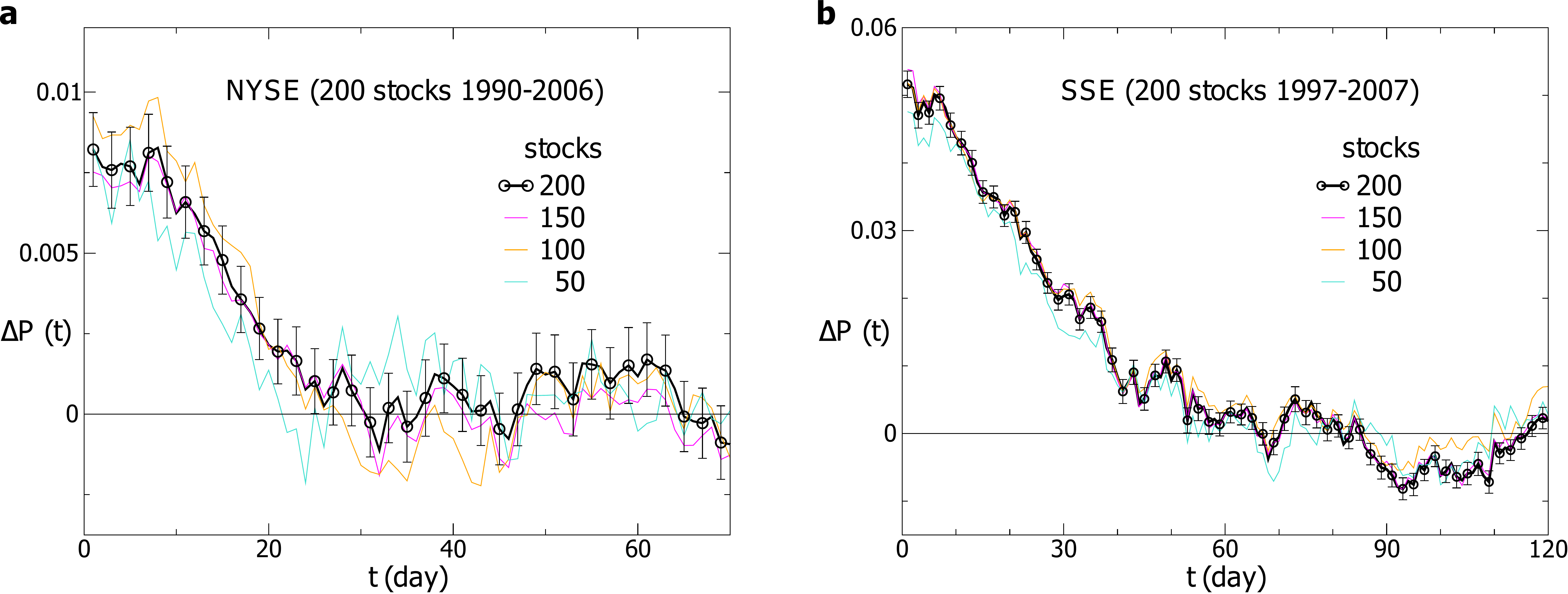}
  \caption{\textbf{The probability difference for the individual stocks}. The probability difference $\Delta P(t)$ for (\textbf{a}) $200$
individual stocks in the SSE and (\textbf{b}) $200$ individual stocks in the NYSE. The black line shows $\Delta P(t)$ averaged over
$200$ stocks with error bars. The other lines represent $\Delta P(t)$ averaged over $150$, $100$, and $50$ randomly chosen stocks.
The time windows are $T_{1}=24$ and $T_{2}=205$ for the NYSE, and $T_{1}=10$ and $T_{2}=210$ for the SSE.
   }
  \label{Fig1}
\end{figure}

At the time windows $T_{1}=24$ and $T_{2}=205$, for instance, we compute $\Delta P(t)$ for $200$ stocks
in the NYSE and take an average over these stocks.
As displayed in Fig.~\ref{Fig1}(a), the average $\Delta P(t)$ remains positive for over $20$ days with an amplitude of $1$ percent.
The result indicates that the past volatilities nonlocal in time enhance the positive returns in the future.
For comparison, three curves for $\Delta P(t)$ averaged over $150$, $100$ and $50$ randomly chosen stocks are also displayed. Within
fluctuations, these three curves are consistent with that for $\Delta P(t)$ averaged over $200$ stocks.
We take the average over many stocks for the purpose of exploring the collective behavior of stocks.
For a single stock, the price dynamics is much more complicated, and $\Delta P(t)$ fluctuates more strongly.
Then we perform the same
computation for $200$ stocks in the SSE at the time windows $T_{1}=10$ and $T_{2}=210$. As displayed in
Fig.~\ref{Fig1}(b), $\Delta P(t)$ remains positive for about $40$ days and the amplitude is about $5$ percent.
Compared with the results for the NYSE, the amplitude and duration of $\Delta P(t)$ for the SSE are respectively much larger and longer.
The reason may be that the US stock market is highly developed, with large market size and diversified investment philosophies,
while the Chinese stock market is emerging and of small market size, in which the investment philosophies of investors resemble each other.

For the validation of our methods, each point of $\Delta P(t)$ in Fig.~\ref{Fig1} is analyzed by performing Student's $t$-test.
In general, a $p$-value less than $0.01$ is considered statistically significant. For the NYSE, the smallest $p$-value is in the order of $10^{-12}$,
and all the $p$-values for $1\leqslant t\leqslant19$ are less than $0.01$. For the SSE, the $p$-values are even smaller, and less than $0.01$ for
$1\leqslant t\leqslant52$.

Actually the definition of volatility in Eq.~(\ref{e20}) is a simplified one. A more standard definition of volatility at $t'$ is
\begin{equation}
v_{1}\left(t'\right)=\left[\frac{1}{m}\sum_{i=1}^{m}r^{2}\left(t'-i+1\right)\right]^{1/2},
\label{e60}
\end{equation}
where $m$ represents a relatively small time window, which
may be set to be $5$ days, i.e., the number of the trading days in a week. Given that these two definitions $v(t')$ and $v_1(t')$ may lead to different results
in extreme volatility regimes, we consider both of them in our calculations.
For $v_{1}\left(t'\right)$, the average volatility at $t'$ over a past period of time $T$ is
$\left\langle v_{1}(t')\right\rangle _{T}=[\sum_{i=1}^{T-m+1}v_{1}\left(t'-i+1\right)]/\left(T-m+1\right)$, with $T\geqslant m$.
Thus, the difference of the average volatilities in two different time windows is
$\Delta v_{1}\left(t'\right)=\left\langle v_{1}\left(t'\right)\right\rangle _{T_{1}}-\left\langle v_{1}\left(t'\right)\right\rangle _{T_{2}}$.

For further comparison, one may also define the average volatility at $t'$ over a past period of time $T$ as
$\left\langle v_{2}(t')\right\rangle _{T}=[1/T\cdot\sum_{i=1}^{T}r^{2}(t'-i+1)]^{1/2}$. Thus the difference of the average volatilities in two different
time windows is $\Delta v_{2}\left(t'\right)=\left\langle v_{2}\left(t'\right)\right\rangle _{T_{1}}-\left\langle v_{2}\left(t'\right)\right\rangle _{T_{2}}$.
For $\Delta v_{1}\left(t'\right)$ and $\Delta v_{2}\left(t'\right)$ respectively, the probability difference is
\begin{equation}
\Delta P_{1}(t)=P^+(t)|_{\Delta v_{1}(t')>0}-P^+(t)|_{\Delta v_{1}(t')<0}
\end{equation}
and
\begin{equation}
\Delta P_{2}(t)=P^+(t)|_{\Delta v_{2}(t')>0}-P^+(t)|_{\Delta v_{2}(t')<0}.
\end{equation}

\begin{figure}[H]
 \centering
    \includegraphics[height=2.2in]{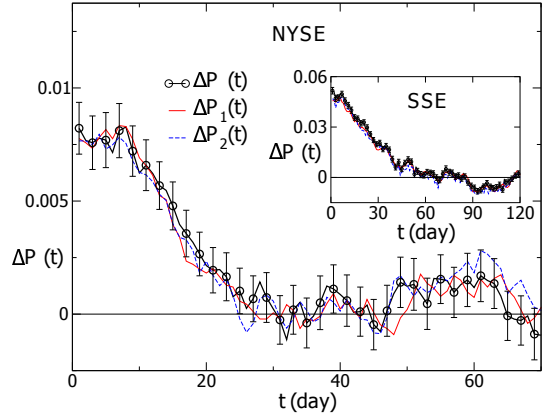}
  \caption{\textbf{The probability differences for three definitions of volatility.} The results are averaged over individual stocks. For the NYSE,
the time windows are $T_{1}=24$ and $T_{2}=205$. The result for the SSE is displayed in the inset, with the time windows $T_{1}=10$ and $T_{2}=210$.
   }
  \label{Fig2}
\end{figure}

For the NYSE and SSE respectively,we compute $\Delta P_{1}(t)$ and $\Delta P_{2}(t)$, and take an average over individual stocks.
The time windows are the same as those for $\Delta P(t)$ in Fig.~\ref{Fig1}. As displayed in Fig.~\ref{Fig2}, the curves for $\Delta P(t)$, $\Delta P_{1}(t)$
and $\Delta P_{2}(t)$ overlap each other within fluctuations. In the following calculations, we mainly consider $\Delta P(t)$ and $\Delta P_{1}(t)$.

\subsection*{Effective pairs of $T_{1}$ and $T_{2}$}
In the calculation of $\Delta P(t)$, the time windows $T_{1}$ and $T_{2}$ are crucial. $T_{1}$ represents the recent period of time, and investors
measure the current fluctuation of prices according to the volatility averaged over $T_{1}$. Thus, $T_{1}$ should be relatively small. In our calculations,
$T_{1}$ ranges from $1$ to $44$ days. Here $44$ is the number of trading days in two months. $T_{2}$ stands for
the period of time in which one estimates the background of volatilities in the past. Theoretically, $T_{2}$ should be much larger than $T_{1}$.
On the other hand, $T_{2}$ should not be arbitrarily large either: firstly, the memory of
investors may not last very long; secondly, maybe more importantly,
$T_{2}$ reflects the long-term fluctuation of stock markets, which should be reasonably fixed.
In our calculations, $T_{2}$ ranges from $45$ to $250$ days. Here $250$ is the number of trading days in a year.
In fact, $T_{2}$ is more crucial to $\Delta P(t)$ than $T_{1}$. If $T_{2}$ were equal to the total length of the volatility series,
$\left\langle v\left(t'\right)\right\rangle _{T_{2}}$ would be a constant, and $\Delta P(t)$ would become a local observable,
which is just a volatility-return correlation function local in time but averaged over a $T_{1}$-day moving window.

In Fig.~\ref{Fig1}(a) and (b), we display $\Delta P(t)$ computed with a specific effective pair of $T_{1}$ and $T_{2}$. Actually, the effective pair of
$T_{1}$ and $T_{2}$ is not unique, and exists in a particular region.
Therefore, we compute $\Delta P(t)$ with each pair of $T_{1}$ and $T_{2}$, and identify the effective
pairs at which $\Delta P(t)$ is significantly non-zero. Since $\Delta P(t)$ needs to be computed in a large region of $T_{1}$ and $T_{2}$,
it is inefficient to observe the behavior of $\Delta P(t)$ by eyes. Besides, due to the fluctuation of $\Delta P(t)$, the visual observation
could be difficult in some cases. Therefore, we propose technical criteria to efficiently discriminate the non-zero $\Delta P(t)$.

The schematic diagram of the criteria is displayed in Fig.~\ref{Fig3}. The criteria comprise four steps:

\begin{figure}[H]
 \centering
    \includegraphics[height=2.2in]{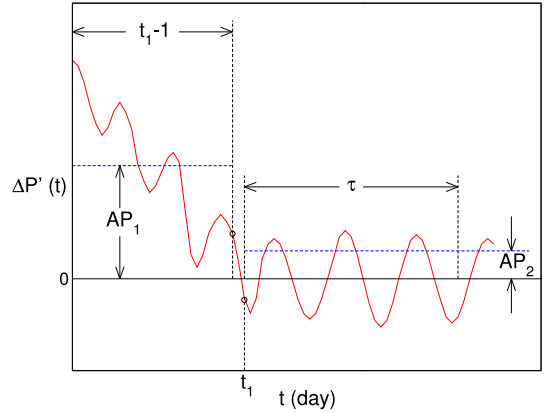}
  \caption{\textbf{A schematic graph of the criteria for identifying non-zero $\Delta P$.} The red line represents $\Delta P'(t)$,which is the 3-point smoothed
$\Delta P(t)$. $t_{1}$ is the day when the sign of $\Delta P'(t)$ changes for the first time. We define $\Delta P'(t)$ in the range of $1\leq t\leq t_{1}-1$ as
the first part, and that in the range of $t_{1}\leq t\leq t_{1}+\tau-1$ as the second part. The average absolute values for the first and second parts are denoted
by $AP_{1}$ and $AP_{2}$, respectively.
   }
  \label{Fig3}
\end{figure}

(1)  $\Delta P(t)$ is smoothed with a 3-day moving window and the result is denoted by $\Delta P'(t)$.

(2) Supposing $\Delta P'(t)$ changes sign for the first time at $t_{1}$, we define $\Delta P'(t)$ in the range of $1\leq t\leq t_{1}-1$ as the first part,
and that in the range of $t_{1}\leq t\leq t_{1}-1+\tau$ as the second part. A non-zero $\Delta P'(t)$ would remain positive or negative in the
first part, while fluctuate around zero in the second part. We set $\tau$ to be $44$, i.e., the number of the trading days in two months, which is long
enough to confirm whether the second part of $\Delta P'(t)$ fluctuates around zero.

(3) we calculate the average absolute values for the first and second parts of $\Delta P'(t)$, denoted by $AP_{1}$ and $AP_{2}$ respectively.

(4) For a non-zero $\Delta P(t)$, it has to be satisfied that
(i)$\Delta P'(t)>AP_{2}$ for $1\leqslant t\leqslant t_{0}$, and $t_{0}>10$;
(ii) each value of $\Delta P'(t)$ in the second part is smaller than $AP_{1}$.
With these conditions, we sift out the non-zero $\Delta P(t)$ preliminarily. To measure how significantly $\Delta P(t)$ differs from zero, we calculate
the average value of $\Delta P'(t)$ for $1\leqslant t\leqslant t_{0}$, which is denoted by $AP_{0}$.
Actually, the larger $|AP_{0}|$ is, the more significantly $\Delta P(t)$ differs from zero. The average of non-zero $AP_{0}$ over different pairs of
$T_{1}$ and $T_{2}$ is denoted by $\overline{AP_{0}}$. To consolidate our results, we identify those non-zero $\Delta P(t)$, which meet
an additional requirement:
(iii) $|AP_{0}|>|\overline{AP_{0}}|$.
$AP_{0}$ is set to $0$ unless $\Delta P(t)$ satisfies all the requirements above.

\begin{figure}[H]
 \centering
    \includegraphics[height=1.9in]{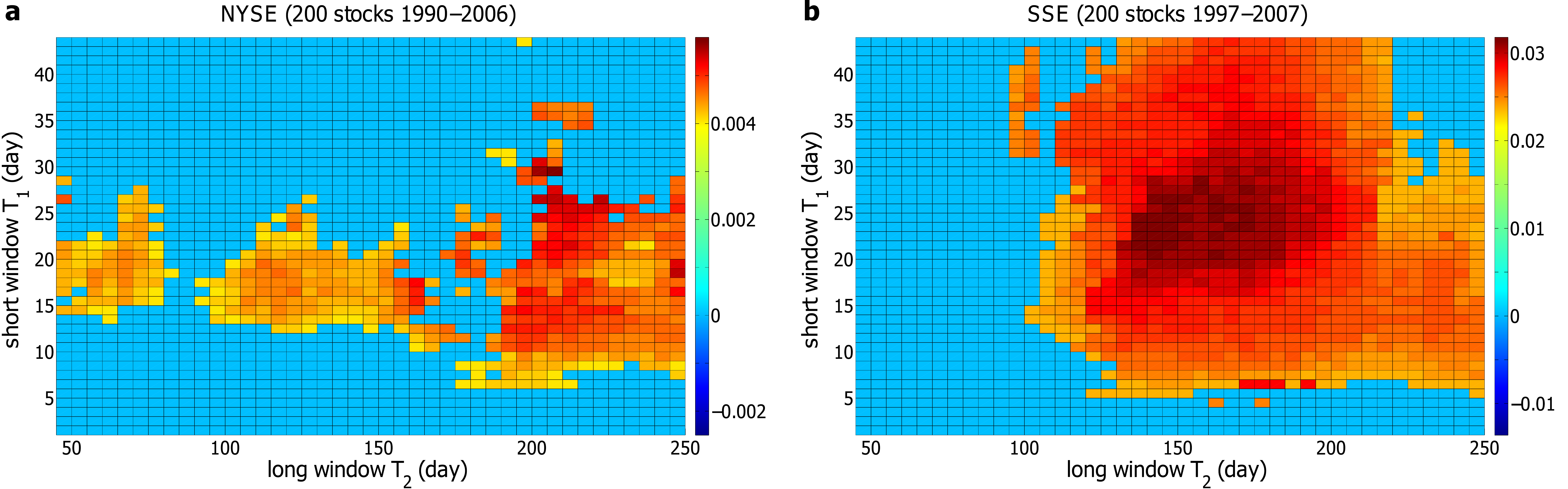}
  \caption{\textbf{The landscape for the amplitude of $\Delta P(t)$.} The amplitude $AP_{0}$ of $\Delta P(t)$ at different time windows
$T_{1}$ and $T_{2}$ for (\textbf{a}) $200$ individual stocks in the NYSE and (\textbf{b}) $200$ individual stocks in the SSE. $T_{1}$ ranges
from $1$ day to $44$ days, and the increment is $1$ day. $T_{2}$ is from $45$ to $250$ days, with an increment of $5$ days. The larger
$AP_{0}$ is, the more significantly $\Delta P(t)$ differs from zero. For $\Delta P(t)$ fluctuating around zero, $AP_{0}=0$.
   }
  \label{Fig4}
\end{figure}

Now we compute $\Delta P(t)$ for the individual stocks in the NYSE with each pair of $T_{1}$ and $T_{2}$. $\Delta P(t)$ is averaged over $200$
stocks, and the corresponding $AP_{0}$ is calculated. The landscape of $AP_{0}$ is displayed in Fig.~\ref{Fig4}(a).
The result indicates that the effective pairs of $T_{1}$ and $T_{2}$ do exist
in a particular region, and both $T_{1}$ and $T_{2}$ are characteristics of the stock markets. In Fig.~\ref{Fig4}(a), the effective
pairs of $T_{1}$ and $T_{2}$ are basically adjacent to each other, suggesting that $\Delta P(t)$ locally is not very sensitive to $T_{1}$ and $T_{2}$.
This is somehow expected, since $\Delta P(t)$ is computed from the volatilities averaged over $T_{1}$ and $T_{2}$,
and a little alteration in $T_{1}$ or $T_{2}$ would not dramatically change $\Delta P(t)$. From this perspective, the gaps between
the disconnected regions in Fig.~\ref{Fig4}(a) probably result from the fluctuations,
especially taking into account the relatively small amplitude of non-zero $\Delta P(t)$ for the NYSE.

Next, we perform a parallel analysis on the $200$ stocks in the SSE, and the landscape for the amplitude of $\Delta P(t)$ is displayed in Fig.~\ref{Fig4}(b).
Similar with the result for the NYSE, a large region of non-zero $\Delta P(t)$ is observed for the SSE. At a single pair of $T_{1}$ and $T_{2}$,
$\Delta P(t)$ averaged over $200$ stocks would generally be non-zero, if $\Delta P(t)$ of some stocks is non-zero.
Moreover, the region
of non-zero $\Delta P(t)$ varies from one stock to another. Therefore, the average $\Delta P(t)$ of the individual stocks is non-zero in a relatively
large region for both the NYSE and SSE. Additionally, as displayed in Fig.~\ref{Fig4}(b), there exists only one connected region
of non-zero $\Delta P(t)$
for the SSE, with the amplitude dwindling from the center to the edge. Compared with the result for the NYSE in Fig.~\ref{Fig4}(a), the region of
non-zero $AP_{0}$ in Fig.~\ref{Fig4}(b) is broader, without gaps, and the value of $AP_{0}$ is almost an order of magnitude larger. The reason
may be traced back to the fact that the Chinese stock market is emerging, and less efficient than the US stock market.
To further validate our methods, we perform Student's $t$-test on each point of non-zero $\Delta P(t)$ in Fig.~\ref{Fig4}.
A $p$-value less than $0.01$ is considered statistically significant. At an effective pair of $T_{1}$ and $T_{2}$, $\Delta P(t)$ is confirmed to be non-zero,
if all the $p$-values are less than $0.01$ for $1\leqslant t\leqslant10$. All non-zero $\Delta P(t)$ are confirmed except for a few ones
at very small $T_{1}$.

We also
compute $\Delta P_{1}(t)$ with different pairs of  $T_{1}$ and $T_{2}$ for the NYSE and SSE. Since $m$ in Eq.~(\ref{e60}) is set to be $5$, $T_{1}$
should not be smaller than $5$. The landscapes for the amplitude of $\Delta P_{1}(t)$ are almost the same as those for the amplitude
of $\Delta P(t)$.

\begin{figure}[H]
 \centering
    \includegraphics[height=1.9in]{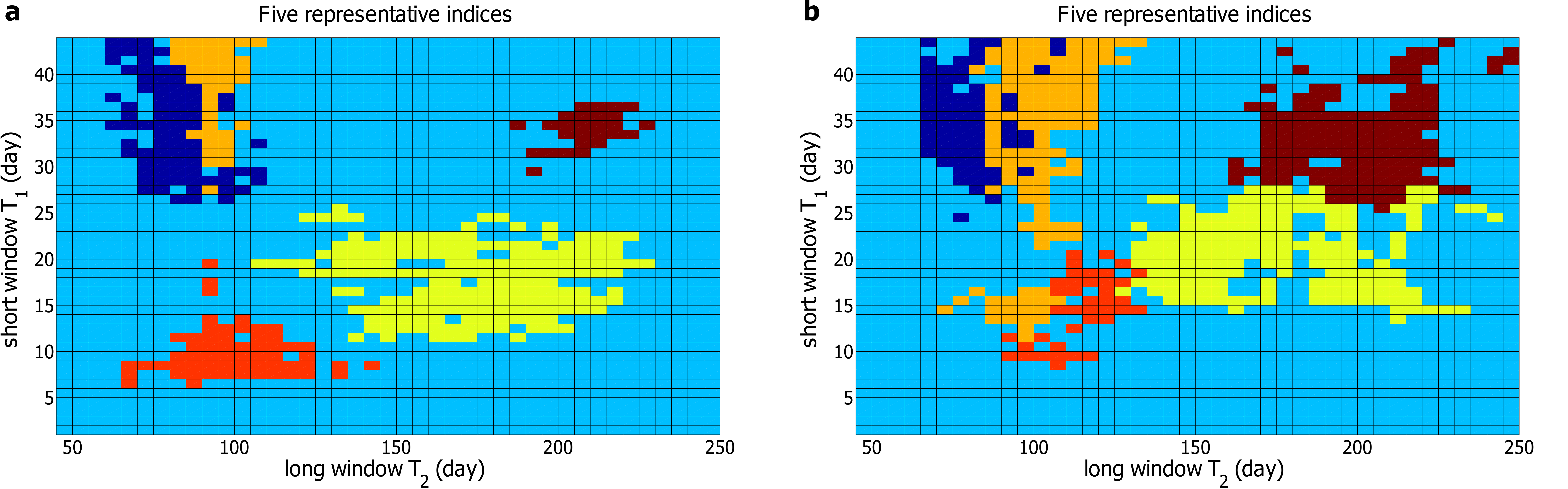}
  \caption{\textbf{The effective pairs of time windows for five representative indices.} The probability difference (\textbf{a}) $\Delta P(t)$
and (\textbf{b}) $\Delta P_{1}(t)$ for five stock market indices. Different colors represent the regions of effective pairs of
$T_{1}$ and $T_{2}$ for different indices. Specifically, navy stands for the Brazil Index, orange stands for the Shanghai Index, red stands for
the Mexico Index, yellow stands for the Spanish Index and crimson stands for the S\&P 500 Index. For clarity, we display only one color
at the overlapping regions, given that these regions are small. Some scattered points are also omitted.
   }
  \label{Fig5}
\end{figure}

Further, we compute $\Delta P(t)$ for the $25$ stock market indices in different countries.
The volatility-return correlation is positive for $16$ indices, and the corresponding effective pairs of $T_{1}$ and $T_{2}$, as well as the maximum
$AP_{0}$, are given in Table~\ref{t1}. For most of these indices, the maximum $AP_{0}$ is over $2$ percent, indicating that the correlation is
rather prominent. In Fig.~\ref{Fig5}(a), we display the regions of effective pairs of $T_{1}$ and $T_{2}$ for $5$ representative indices
including the Brazil, Shanghai, Mexico, Spanish and S\&P 500 indices. For other $7$ indices, nonzero $\Delta P(t)$ could not be detected for
almost all pairs of $T_{1}$ and $T_{2}$. Exceptionally, the Australia and Japan indices exhibit a negative volatility-return correlation, i.e.,
the volatilities in a past period of time enhance the negative returns in future times. The effective pairs of $T_{1}$ and $T_{2}$, as well as the maximum
$AP_{0}$ for these two indices, are also presented in Table~\ref{t1}. We also compute $\Delta P_{1}(t)$ for the $5$
representative indices, and the regions of effective pairs of $T_{1}$ and $T_{2}$ are shown in Fig.~\ref{Fig5}(b). Compared with
Fig.~\ref{Fig5}(a), the regions of the effective pairs of $T_{1}$ and $T_{2}$ in Fig.~\ref{Fig5}(b) change slightly. The reason may be that the fluctuation
of $\Delta P(t)$ and $\Delta P_{1}(t)$ for indices is stronger than that for the individual stocks.

To confirm that the nonlocal volatility-return correlation is indeed a nontrivial dynamic property of the stock markets, we randomly shuffle
the time series of returns, i.e., randomize the time order of the returns, and perform the same calculation.
In this case, $\Delta P(t)$ just fluctuates
around zero. The result provides evidence that the correlation does originate from the interactions between past volatilities and future returns.

\subsection*{Volatility-return correlation functions nonlocal in time}
Up to now, we have only concerned with the signs of $\Delta v(t')$ and $r(t'+t)$ in computing $\Delta P(t)$. Actually, the magnitudes
of $\Delta v(t')$ and $r(t'+t)$ should also be important to both theoretical analysis and practical applications. Taking into account the magnitudes
of $\Delta v(t')$ and $r(t'+t)$, we may explicitly construct a correlation function nonlocal in time to describe the volatility-return correlations,
\begin{equation}
F(t)=\left\langle \Delta v(t')\cdot r(t'+t)\right\rangle.
\label{e80}
\end{equation}
Both $\Delta P(t)$ and $F(t)$ reflect the asymmetric behavior of $r(t'+t)$ in volatile and stable markets, but $\Delta P(t)$ should be more fundamental.
When $\Delta P(t)$ is non-zero, $F(t)$ would be zero only if the contributions of $r(t'+t)>0$ and $r(t'+t)<0$ happen to cancel each other.

\begin{figure}[H]
 \centering
    \includegraphics[height=1.9in]{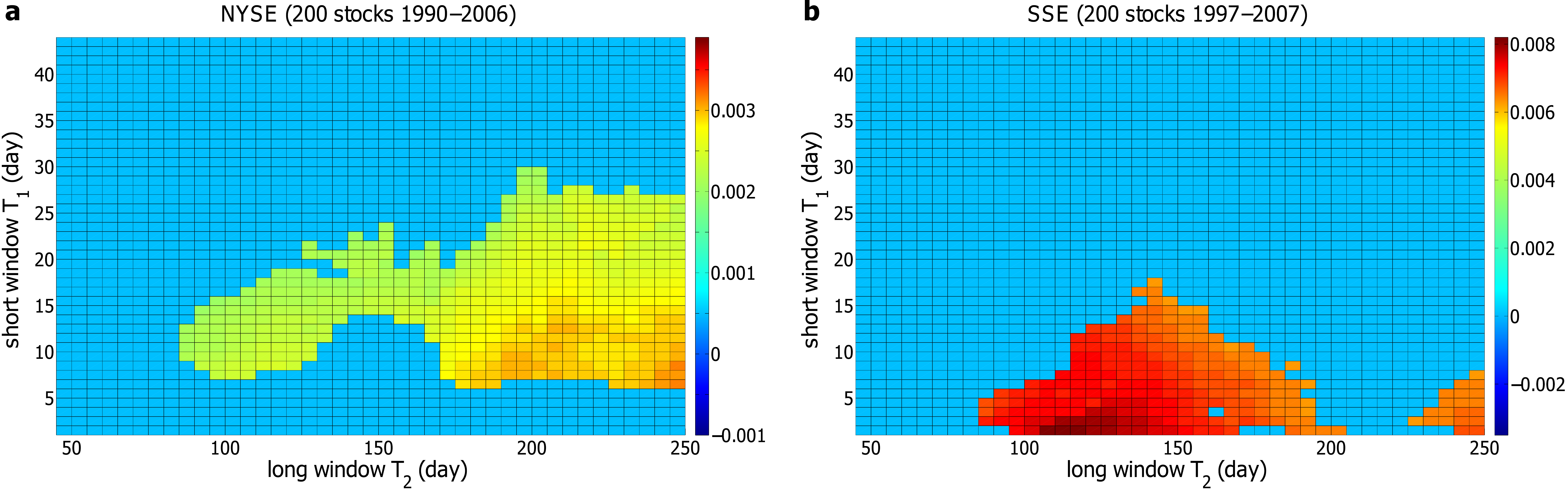}
  \caption{\textbf{The landscape for the amplitude of $F(t)$.} The amplitude $AF_{0}$ of $F(t)$ at different time windows $T_{1}$ and $T_{2}$
for (\textbf{a}) $200$ individual stocks in the NYSE and (\textbf{b}) $200$ individual stocks in the SSE.
   }
  \label{Fig6}
\end{figure}

We compute $F(t)$ with different pairs of $T_{1}$ and $T_{2}$ for the $200$ stocks in the NYSE and SSE respectively,
and identify the non-zero ones with the same criteria for the non-zero $\Delta P(t)$. We also introduce $AF_{0}$ to describe how significantly $F(t)$ differs
from zero, of which the definition is the same as $AP_{0}$ for $\Delta P(t)$. $F(t)$ is averaged over $200$ stocks, and the landscape of the corresponding
$AF_{0}$ is displayed in Fig.~\ref{Fig6}. The dynamic behavior of $F(t)$ is qualitatively the same as that of $\Delta P(t)$ but quantitatively different.
Both the amplitude of $F(t)$ and the region of effective pairs of $T_{1}$ and $T_{2}$ are smaller than those of $\Delta P(t)$. The fluctuation of $F(t)$
is also somewhat stronger.
Student's $t$-test is performed on the non-zero $F(t)$ and almost all of them are confirmed to be non-zero.
We also compute $F_{1}(t)$, which is defined as $F_{1}(t)=\left\langle \Delta v_{1}(t')\cdot r(t'+t)\right\rangle$, with
each pair of $T_{1}$ and $T_{2}$ for the NYSE and SSE, and the results are almost the same as those for $F(t)$.

In fact, $\Delta P(t)$ can be expressed as the correlation function $G(t)=\left\langle sgn(\Delta v(t'))\cdot sgn(r(t'+t))\right\rangle$. Here
$sgn(x)$ represents the sign of $x$. $G(t)$ behaves almost the same way as
$\Delta P(t)$ does. Additionally, one may also define another volatility-return correlation function $H(t)=\left\langle sgn(\Delta v(t'))\cdot r(t'+t)\right\rangle$.
Since only the magnitude of $r(t'+t)$ is taken into consideration, $H(t)$ is less fluctuating than $F(t)$, whereas the result looks qualitatively similar.

There have been many researches with different methods focusing on how volatilities affect returns in financial markets. A direct way is to calculate
the usual volatility-return correlation function, which is defined as $f(t)=\langle v\left(t'\right)\cdot r\left(t'+t\right)\rangle$ with $t>0$. However, the result fluctuates
around zero\cite{bou01,qiu06}. In the past years, various GARCH-like models are applied to investigate the correlation
between past volatilities and future returns. In these models, the future returns are assumed to be correlated with the past volatilities, and there are coefficients
quantifying the correlation. The results are controversial. The correlation is discovered to be positive in some researches\cite{fre87,cam92},
but negative in others\cite{tur89,nel91,glo93}. More often, the coefficient linking past volatilities and future returns is statistically insignificant\cite{bek00}.
From our perspective, these studies only characterize the volatility-return correlation local in time. In our work, however, both $\Delta P(t)$ and $F(t)$
are nonlocal in time,  which are constructed based on the difference between the average volatilities in two different time windows. The correlation
characterized by $\Delta P(t)$ and $F(t)$ is more complicated and of higher-order.

\subsection*{Agent-based model with asymmetric trading preference}
We construct an agent-based model to investigate the microscopic origin of the nonlocal volatility-return correlation.
Agent-based modeling is a promising approach in complex systems, and has been applied successfully to study the fundamental properties in financial markets,
such as the fat-tail distribution of returns, the long-range temporal correlation of volatilities, and the leverage and anti-leverage effects\cite{lux99,con00,egu00,bon02,hom02,sam07,far09,fen12,che13,sor14}.

The basic structure of our model is borrowed from the models in  refs.~\cite{fen12,che13}, which is built on agents' daily
trading, i.e., buying, selling and holding stocks.  Since the information for investors is highly incomplete, an agent's decision of buying, selling or
holding is assumed to be random. Due to the lack of persistent intraday trading in the empirical trading data, we consider that only one trading
decision is made by each agent in a single day.  In our model, there are $N$ agents and each agent only operates one share of stock each day.
On day~$t$, we denote the trading decision of agent~$i$ by
\begin{equation}
S_{i}(t)=\begin{cases}
\:1 & \textnormal{buy}\\
-1 & \textnormal{sell}\\
\:0 & \textnormal{hold}
\end{cases}.
\end{equation}
The probability of buying, selling and holding decisions are denoted by $P_{buy}$, $P_{sell}$ and $P_{hold}$, respectively.
Assuming that the price is determined by the difference between the demand and supply of the stock, we define the return $R(t)$ as
\begin{equation}
R(t)=\sum_{i=1}^{N}S_{i}(t).
\end{equation}

Next, we introduce the investment horizon based on the fact that investors make decisions according to the previous market performance
of different time horizons. It is found that the relative portion $\gamma_{i}$ of investors with $i$ days investment horizon
follows a power-law decay $\gamma_{i}\varpropto i^{- \eta}$ with $\eta=1.12$. With the condition of $\sum_{i=1}^{M}\gamma_{i}=1$,
$\gamma_{i}$ is normalized to be $\gamma_{i}=i^{- \eta}/\sum_{i=1}^{M}i^{- \eta}$, where $M$ is the maximum investment horizon.
Considering different investment horizons of various agents, we introduce a weighted average return $R'(t)$
to describe the integrated investment basis of all agents. Specifically, $R'(t)$ is defined as
\begin{equation}
R'(t)=k\sum_{i=1}^{M}\left[\gamma_{i}\sum_{j=0}^{i-1}R(t-j)\right],
\label{eq:fR}
\end{equation}
where $k$ is a proportional coefficient. We set $k=1/(\sum_{i=1}^{M}\sum_{j=i}^{M}\gamma_{j})$, so that $|R'(t)|_{max}=N=|R(t)|_{max}$.
According to ref.~\cite{che13}, the maximum investment horizon $M$ is set to $150$.

Herding behavior is an important collective behavior in financial markets. We define a herding degree $D(t)$ to describe the clustering degree
of the herding behavior,
\begin{equation}
D(t+1)=|R'(t)|/N.
\label{eq:od}
\end{equation}
On day $t+1$, the average number of agents in each group is $N\cdot D(t+1)$, and therefore we divide all agents into $1/D(t+1)$ groups. The agents
in a same group make a same trading decision with the same trading probability.
In ref.~\cite{fen12}, it is assumed that the probabilities of buying and selling are equal, i.e., $P_{buy}=P_{sell}=p$, and $p$ is a constant estimated to be
$0.0154$. Therefore the trading probability is $P_{trade}=P_{buy}+P_{sell}=2p$ and the holding probability is $P_{hold}=1-2p$. In our model,
the trading probability is also kept to be $2p$ and remains constant during the dynamic evolution.

Now we introduce a novel mechanism in our model, that is, the asymmetric trading preference in volatile and stable markets.
In financial markets, the market behaviors of buying and selling are not always in balance\cite{ple03}.
Hence, $P_{buy}$ and $P_{sell}$ are not always equal to each other. They are affected by previous volatilities,
and the more volatile the market is, the more $P_{buy}$ differs from $P_{sell}$.

For an agent with $i$ days investment horizon, the average volatility over previous $i$ days is taken into account, which is defined as
\begin{equation}
V_{i}(t)=\frac{1}{i}\sum_{j=1}^{i}V(t-j+1).
\end{equation}
Then we define the background volatility as $V_{M}(t)$, where $M$ is the maximum investment horizon. On day $t$, the agent with $i$ days
investment horizon estimates the volatility of the market by comparing $V_{i}(t)$ with $V_{M}(t)$. Therefore, the integrated perspective of all agents
on the recent market volatility is defined as
\begin{equation}
\xi(t)=\left[\sum_{i=1}^{M}\gamma_{i}V_{i}(t)\right]/V_{M}(t).
\end{equation}
Thus, we define the probabilities of buying and selling as
\begin{equation}
\left\{ \begin{array}{l}
P_{buy}(t+1)=p[c\cdot\xi(t)+(1-c)]\\
P_{sell}(t+1)=2p-P_{buy}(t+1)
\end{array}\right..
\label{eq:fp}
\end{equation}
Here the parameter $c$ measures the degree of agents' asymmetric trading preference in volatile and stable markets. Compared with the model
in ref.~\cite{che13}, $c$ is the only new parameter added in our model. We speculate that $c$ can be determined from the trade and quote
data of stock markets. Unfortunately, the data are currently not available to us.

To judge from the amplitude of the volatility-return correlation, $c$ should be a small number. Let us set $c$ to be $1/80$.
The total number of the agents, $N$, is $10000$. The returns of the initial $150$
time steps are set to be random values following a standard Gaussian distribution. On day $t$, we randomly divide $N$ agents into $1/D(t)$ groups.
The agents in a same group make a same trading decision with the same probability. After each agent makes his decision, the return $R(t)$ can be computed.
Repeating the procedure we produce $20000$ data points of $R(t)$ in each simulation, and abandon the first $15000$ data points for
equilibration. Thus we obtain a sample with 5000 data points.

\begin{figure}[H]
 \centering
    \includegraphics[height=1.9in]{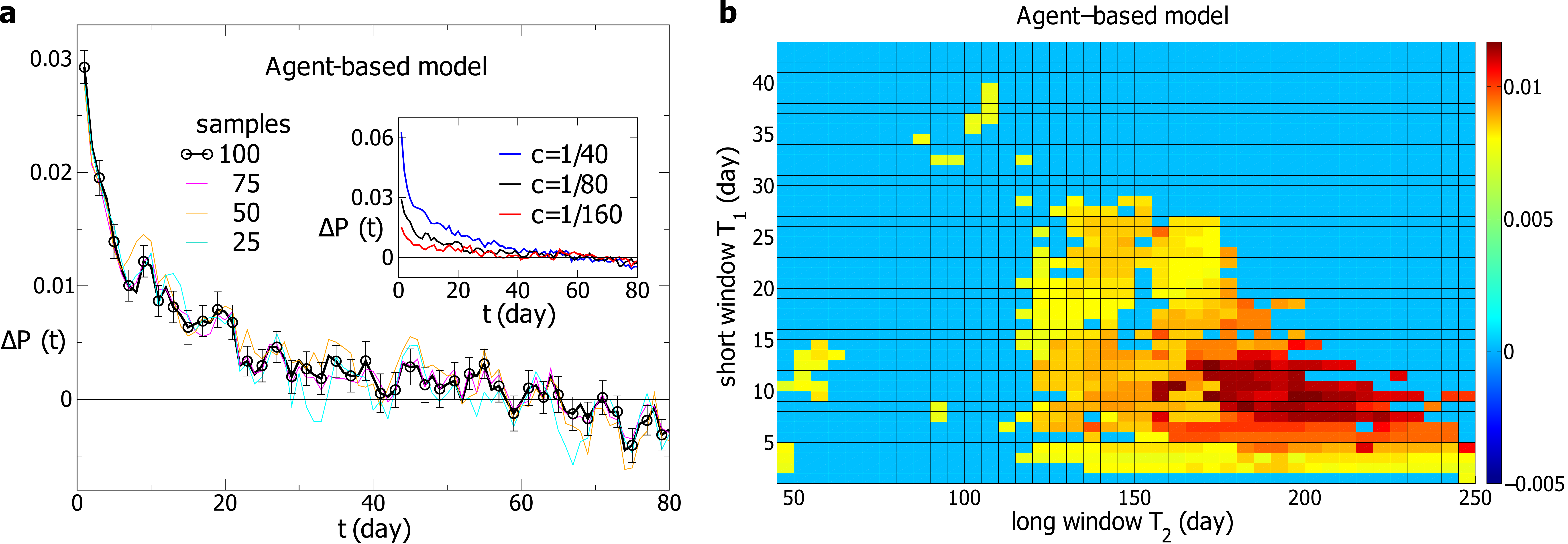}
  \caption{\textbf{The simulation results of the agent-based model.} (\textbf{a}) The probability difference $\Delta P(t)$ computed with the simulated returns
at the time windows $T_{1} = 3$ and $T_{2} = 150$. The parameter $c$ is $1/80$. The black line represents $\Delta P(t)$ averaged over $100$ samples
with error bars and the other lines stand for $\Delta P(t)$ averaged over $75$, $50$ and $25$ randomly chosen samples.
$\Delta P(t)$ for different values of $c$ are displayed in the inset, with $T_{1} = 3$ and $T_{2} = 150$. (\textbf{b}) The amplitude
$AP_{0}$ of $\Delta P(t)$ at different time windows $T_{1}$ and $T_{2}$. Each $\Delta P(t)$ is averaged over $100$ samples.
The larger $AP_{0}$ is, the more significantly $\Delta P(t)$ differs from zero. For $\Delta P(t)$ fluctuating around zero, $AP_{0}=0$.
   }
  \label{Fig7}
\end{figure}

After the time series $R(t)$ generated from our model is normalized to $r(t)$, we compute $\Delta P(t)$ with the time windows $T_{1}=3$ and $T_{2}=150$.
The result is averaged over $100$ samples and displayed in Fig.~\ref{Fig7}(a).  $\Delta P(t)$ is significantly non-zero
with an amplitude of $3$ percent, lasting for about $20$ days. For comparison, three curves for $\Delta P(t)$ averaged over $75$, $50$ and $25$
randomly chosen samples are also displayed. Within fluctuations, these four curves are consistent with each other
and in agreement with the empirical results.

We also perform the simulation with $c=1/40$ and $c=1/160$, respectively, to investigate the dependence of $\Delta P(t)$ on $c$.
As displayed in the inset of Fig.~\ref{Fig7}(a), the amplitude of $\Delta P(t)$ increases with $c$, i.e., the magnitude of $c$ determines the amplitude of the volatility-return
correlation. For $c=1/40$, the amplitude of $\Delta P(t)$ is about $6$ percent, which is in the order of that for the SSE and other markets with a strong volatility-return
correlation. For $c=1/160$, the amplitude of $\Delta P(t)$ is close to that for the S\&P500 index, of which the volatility-return
correlation is relatively weak. Therefore, with $c$ ranging from $1/160$ to $1/40$, our model produces the volatility-return correlation consistent with the empirical results.
Additionally, if $c$ is negative,
the volatility-return correlation will be negative, i.e., the sign of $c$ fixes the correlation to be positive or negative.

Next, we compute $\Delta P(t)$ with different pairs of $T_{1}$ and $T_{2}$, and determine the region of effective pairs of $T_{1}$ and $T_{2}$.
$\Delta P(t)$ is averaged over $100$ samples, and the landscape of $AP_{0}$ is shown in Fig.~\ref{Fig7}(b). A single region
with non-zero $\Delta P(t)$ is observed. For $T_{2}$ smaller than $120$, for example, $\Delta P(t)$ is almost zero. In other words,
the effective pairs of $T_{1}$ and $T_{2}$ exist in a particular region, which is consistent with the empirical results.

\section*{Discussion}
We construct a class of dynamic observables nonlocal in time to explore the correlation between past volatilities and future returns in stock
markets. Strikingly, the volatility-return correlation is discovered to be non-zero, with an amplitude of a few percent and
a duration of over two weeks. The result indicates that past volatilities nonlocal in time affect future returns. Both the nonlocal dynamic observables $\Delta P(t)$
and $F(t)$ rely on two time windows  $T_{1}$ and $T_{2}$. The effective pairs of $T_{1}$ and $T_{2}$ exist in a particular region, suggesting that both $T_{1}$
and $T_{2}$ are the characteristics of the stock markets.

Our results are robust for not only individual stocks but also stock market indices. The volatility-return correlation nonlocal in time is detected to be positive for
individual stocks in the New York and Shanghai stock exchanges, as well as $16$ stock indices. For other $7$ indices, $\Delta P(t)$ fluctuates around zero.
However, we suppose there may exist some higher-order correlations between volatilities and returns for these indices, which
could be described by more complicated nonlocal observables. Exceptionally, other $2$ indices exhibit a negative volatility-return correlation.

To investigate the microscopic origin of the volatility-return correlation, we construct an agent-based model with a novel mechanism, that is, the asymmetric
trading preference in volatile and stable markets. Accordingly, a parameter $c$ is introduced to describe the degree of the asymmetric trading preference.
The simulation results exhibit a positive correlation which is in agreement with the empirical ones. More importantly, the effective
pairs of $T_{1}$ and $T_{2}$ for simulation results exist in a particular region, which is also consistent with the empirical ones. Actually, our model can
also produce a negative correlation by changing the sign of $c$. The results reveal that both the positive and negative correlations arise from the
asymmetric trading preference in volatile and stable markets.
In our model, the nonlocality arises from the interaction between the integrated perspective on the recent market volatility and the probabilities of buying and
selling.

Our results provide new insight into the price dynamics. Contrary to the assumptions in various models, the rise and fall of prices
turn out to be far from random. To the best of our knowledge, the volatility-return correlation nonlocal in time is the only property concerning the control
of the price dynamics, given that the autocorrelating time of returns is extremely short. This non-zero volatility-return correlation implies that there may exist
higher-order correlations of returns,  which deserves further investigation in the future, especially for those $7$ indices with
$\Delta P(t)$ fluctuating around zero. Furthermore, our results indicate that nonlocality is an intrinsic characteristic
in the financial markets, which is more important than we thought before. Besides the volatility-return correlation in the stock markets, many other nonlocal
correlations in financial systems are to be explored, which serves as our future agenda.

\section*{Supporting Information}

\subsection*{S1 Appendix}
\label{S1_Appendix}
{\bf Calculation for $P^+(t)|_{\Delta v(t')>0}$, $P^+(t)|_{\Delta v(t')<0}$ and $P_{0}(t)$, and relation among them.}


\section*{Acknowledgments}


\begin{thebibliography}{10}

\bibitem{man95}
Mantegna RN, Stanley HE. Scaling behavior in the dynamics of an economic index. Nature. 1995;376: 46.

\bibitem{ple99}
Plerou V, Gopikrishnan P, Rosenow B, Amaral LAN, Stanley HE. Universal and nonuniversal properties of cross correlations in financial time
series. Phys Rev Lett. 1999;83: 1471.

\bibitem{gop99}
Gopikrishnan P, Plerou V, Amaral LAN, Meyer M, Stanley HE. Scaling of the distribution of fluctuations of financial market indices. Phys Rev E.
1999;60: 5305.

\bibitem{bou01}
Bouchaud JP, Matacz A, Potters M. Leverage effect in financial markets: The retarded volatility model. Phys Rev Lett. 2001;87: 228701.

\bibitem{kra02}
Krawiecki A, Ho{\l}yst JA, Helbing D. Volatility clustering and scaling for financial time series due to attractor bubbling. Phys Rev Lett.
2002;89: 158701.

\bibitem{gab03}
Gabaix X, Gopikrishnan P, Plerou V, Stanley HE. A theory of power-law distributions in financial market fluctuations. Nature. 2003;423: 267.

\bibitem{onn03}
Onnela JP, Chakraborti A, Kaski K, Kertesz J, Kanto A. Dynamics of market correlations: taxonomy and portfolio analysis. Phys Rev E. 2003;68:
056110.

\bibitem{sor03}
Sornette D. Critical market crashes. Phys Rep. 2003;378: 1--98.

\bibitem{qiu06}
Qiu T, Zheng B, Ren F, Trimper S. Return-volatility correlation in financial dynamics. Phys Rev E. 2006;73: 065103.

\bibitem{pod09}
Podobnik B, Horvati´c D, Petersen AM, Stanley HE. Cross-correlations between volume change and price change. Proc Natl Acad Sci USA. 2009;106:
22079.

\bibitem{she09}
Shen J, Zheng B. Cross-correlation in financial dynamics. Europhys Lett. 2009;86: 48005.

\bibitem{ken11}
Kenett DY, Shapira Y, Madi A, Bransburg-Zabary S, Gur-Gershgoren G, Ben-Jacob E. Index cohesive force analysis reveals that the us market became
prone to systemic collapses since 2002. PLoS One. 2011;6: e19378.

\bibitem{pre11}
Preis T, Schneider JJ, Stanley HE. Switching processes in financial markets. Proc Natl Acad Sci USA. 2011;108: 7674--7678.

\bibitem{sha11}
Shapira Y, Kenett DY, Raviv O, Ben-Jacob E. Hidden temporal order unveiled in stock market volatility variance. AIP Advances. 2011;1: 022127.

\bibitem{fen12}
Feng L, Li BW, Podobnik B, Preis T, Stanley HE. Linking agent-based models and stochastic models of financial markets. Proc Natl Acad Sci USA.
2012;109: 8388--8393.

\bibitem{ken12}
Kenett DY, Preis T, Gur-Gershgoren G, Ben-Jacob E. Quantifying meta-correlations in financial markets. Europhys Lett. 2012;99: 38001.

\bibitem{pre12}
Preis T, Kenett DY, Stanley HE, Helbing D, Ben-Jacob E. Quantifying the behavior of stock correlations under market stress. Sci Rep. 2012;2:
752.

\bibitem{che13}
Chen JJ, Zheng B, Tan L. Agent-based model with asymmetric trading and herding for complex financial systems. PLoS One. 2013;8: e79531.

\bibitem{ken13}
Kenett DY, Ben-Jacob E, Stanley HE, Gur-Gershgoren G. How high frequency trading affects a market index. Sci Rep. 2013;3: 2110.

\bibitem{jia14}
Jiang XF, Chen TT, Zheng B. Structure of local interactions in complex financial dynamics. Sci Rep. 2014;4: 5321.

\bibitem{maj14}
Majdandzic A, Podobnik B, Buldyrev SV, Kenett DY, Havlin S, Stanley HE. Spontaneous recovery in dynamical networks. Nat Phys. 2014;10: 34--38.

\bibitem{yur14}
Yura Y, Takayasu H, Sornette D, Takayasu M. Financial brownian particle in the layered order-book fluid and fluctuation-dissipation relations.
Phys Rev Lett. 2014;112: 098703.

\bibitem{bla76}
Black F. Studies of stock price volatility changes. Alexandria: Proceedings of the 1976 Meetings of the American Statistical Association,
Business and Economical Statistics Section. 1976;177--181.

\bibitem{fre87}
French KR, Schwert GW, Stambaugh RF. Expected stock returns and volatility. J financ econ. 1987;19: 3--29.

\bibitem{cam92}
Campbell JY, Hentschel L. No news is good news: An asymmetric model of changing volatility in stock returns. J financ econ. 1992;31: 281--318.

\bibitem{bek00}
Bekaert G, Wu G. Asymmetric volatility and risk in equity markets. Rev Financ Stud. 2000;13: 1--42.

\bibitem{con00}
Cont R, Bouchaud JP. Herd behavior and aggregate fluctuations in financial markets. Macroeconomic Dyn. 2000;4: 170.

\bibitem{yam05}
Yamasaki K, Muchnik L, Havlin S, Bunde A, Stanley HE. Scaling and memory in volatility return intervals in financial markets. Proc Natl Acad Sci
USA. 2005;102: 9424--9428.

\bibitem{bol06}
Bollerslev T, Litvinova J, Tauchen G. Leverage and volatility feedback effects in highfrequency data. J financ econ. 2006;4: 353--384.

\bibitem{osi07}
Osipov GV, Kurths J, Zhou C. Synchronization in oscillatory networks. 1st ed. Berlin: Springer; 2007.

\bibitem{sha09}
Shapira Y, Kenett DY, Ben-Jacob E. The index cohesive effect on stock market correlations. Eur Phys J B. 2009;72: 657--669.

\bibitem{ken10a}
Kenett DY, Shapira Y, Madi A, Bransburg-Zabary S, Gur-Gershgoren G, Ben-Jacob E. Dynamics of stock market correlations. AUCO Czech Economic
Review. 2010;4: 330--341.

\bibitem{ken10b}
Kenett DY, Tumminello M, Madi A, Gur-Gershgoren G, Mantegna RN, Ben-Jacob E. Dominating clasp of the financial sector revealed by partial
correlation analysis of the stock market. PLoS One. 2010;5: e15032.

\bibitem{ren10}
Ren F, Zhou WX. Recurrence interval analysis of high-frequency financial returns and its application to risk estimation. New J Phys. 2010;12:
075030.

\bibitem{da11}
Da Z, Engelberg J, Gao PJ. In search of attention. J Finance. 2011;66: 1461--1499.

\bibitem{zha11}
Zhao L, Yang G, Wang W, Chen Y, Huang JP, Ohashi H, et al. Herd behavior in a complex adaptive system. Proc Natl Acad Sci USA. 2011;108: 15058
--15063.

\bibitem{jia12}
Jiang XF, Zheng B. Anti-correlation and subsector structure in financial systems. Europhys Lett. 2012;97: 48006.

\bibitem{liu99}
Liu Y, Gopikrishnan P, Cizeau P, Meyer M, Peng CK, Stanley HE. Statistical properties of the volatility of price fluctuations. Phys Rev E.
1999;60: 1390.

\bibitem{egu00}
Eguiluz VM, Zimmermann MG. Transmission of information and herd behavior: An application to financial markets. Phys Rev Lett. 2000;85: 5659.

\bibitem{cox76}
Cox JC, Ross SA. The valuation of options for alternative stochastic processes. J financ econ. 1976;3: 145.

\bibitem{she09a}
Shen J, Zheng B. On return-volatility correlation in financial dynamics. Europhys Lett. 2009;88: 28003.

\bibitem{cam88}
Campbell JY, Shiller RJ. The dividend-price ratio and expectations of future dividends and discount factors. Rev Financ Stud. 1988;1: 195--228.

\bibitem{fam88}
Fama EF, French KR. Dividend yields and expected stock returns. J financ econ. 1988;22: 3--25.

\bibitem{val03}
Valkanov R. Long-horizon regressions: theoretical results and applications. J financ econ. 2003;68: 201--232.

\bibitem{bou07}
Boudoukh J, Michaely R, Richardson M, Roberts MR. On the importance of measuring payout yield: Implications for empirical asset pricing. J Finance. 2007;62: 877--915.

\bibitem{moa13}
Moat HS, Curme C, Avakian A, Kenett DY, Stanley HE, Preis T. Quantifying wikipedia usage patterns before stock market moves. Sci Rep. 2013;3:
1801.

\bibitem{pre13}
Preis T, Moat HS, Stanley HE. Quantifying trading behavior in financial markets using google trends. Sci Rep. 2013;3: 1684.

\bibitem{bol86}
Bollerslev T. Generalized autoregressive conditional heteroskedasticity. J Econom. 1986;31: 307--327.

\bibitem{nel91}
Nelson DB. Conditional heteroskedasticity in asset returns: A new approach. Econometrica. 1991;59: 347--370.

\bibitem{tur89}
Turner CM, Startz R, Nelson CR. A markov model of heteroskedasticity, risk, and learning in the stock market. J financ econ. 1989;25: 3--22.

\bibitem{glo93}
Glosten LR, Jaganathan R, Runkle DE. On the relation between the expected value and the volatility of the nominal excess return on stocks. J Finance. 1993;48: 1779--1801.

\bibitem{lux99}
Lux T, Marchesi M. Scaling and criticality in a stochastic multi-agent model of a financial market. Nature. 1999;397: 498.

\bibitem{bon02}
Bonabeau E. Agent-based modeling: Methods and techniques for simulating human systems. Proc Natl Acad Sci USA. 2002;99: 7280--7287.

\bibitem{hom02}
Hommes CH. Modeling the stylized facts in finance through simple nonlinear adaptive systems. Proc Natl Acad Sci USA. 2002;99: 7221--7228.

\bibitem{sam07}
Samanidou E, Zschischang E, Stauffer D, Lux T. Agent-based models of financial markets. Rep Prog Phys. 2007;70: 409.

\bibitem{far09}
Farmer JD, Foley D. The economy needs agent-based modelling. Nature. 2009;460: 685--686.

\bibitem{sor14}
Sornette D. Physics and financial economics (1776--2014): puzzles, ising and agent-based models. Rep Prog Phys. 2014;77: 062001.

\bibitem{ple03}
Plerou V, Gopikrishnan P, Stanley HE. Econophysics: Two-phase behaviour of financial markets. Nature. 2003;421: 130.

\end{thebibliography}
%
%
%

%
%
%

\end{document}